\documentclass{osa-article}

\journal{osac}

\articletype{Research Article}

\usepackage{lineno}

\begin{document}

\title{Design of proton deflectometry with \textit{in situ} X-ray fiducial for magnetized HED systems} 

\author{Sophia Malko\authormark{1,*}, Courtney Johnson\authormark{2}, Derek B. Schaeffer\authormark{2}, William Fox\authormark{1,2}, Gennady Fiksel\authormark{3}}

\address{\authormark{1}Princeton Plasma Physics Laboratory, 100 Stellarator Road, Princeton, NJ 08540\\
\authormark{2}Dept.~of Astrophysical Sciences, Princeton University, Princeton, NJ 08544\\
\authormark{3}Center for Ultrafast Optical Science, University of Michigan, Ann Arbor, MI 48109\\
}
 
\email{\authormark{*}smalko@pppl.gov} 



\begin{abstract}

We report a design and implementation of proton radiography with an \textit{in situ} reference X-ray image of a mesh to precisely measure non-uniform magnetic fields in expanding plasmas at the OMEGA and OMEGA EP laser facilities. The technique has been developed with proton and x-ray sources generated from both directly-driven capsule implosions and short pulse laser-solid interactions. The accuracy of the measurement  depends on the contrast of both the proton and x-ray images. We present numerical and analytic studies to optimize the image contrast using a variety of mesh materials and grid spacing. Our results show a clear enhancement of the image contrast by a factor of 4-6 using a high Z mesh with large grid spacing. This would lead to a further factor of two improvement in accuracy of the magnetic field measurement.
\end{abstract}

\section{Introduction}

The interaction of laser-produced high-energy-density (HED) plasma with a magnetic field plays a key role in magnetized fusion schemes such as MagLIF \cite{Slutz:2010} and laboratory astrophysics experiments such as magnetic reconnection \cite{Fox:2017,Fox:2020} and magnetized collisionless shocks \cite{Schaeffer:2017}. Similar dynamics between strongly-driven flows and magnetic fields are also prevalent in prior space-plasma experiments such as the Active Magnetospheric Particle Tracer Explorers (AMPTE) \cite{Valenzuela:1986}. The understanding of magnetic field transport properties in plasmas is thus a primary goal for these applications, and one aspect of particular interest is the anomalously fast diffusion of plasma caused by strong plasma parameter gradients that lead particles and fields to mix and diffuse with respect to one another. However, the direct measurement of the magnetic field dynamics in such conditions is extremely challenging and requires the development of specific diagnostics. 

One of the most common techniques used for measuring electromagnetic fields in HED plasmas is proton deflectometry (radiography), in which protons passing through the electromagnetic fields in a plasma are deflected and then recorded on a detector \cite{Kugland:2012}. By measuring the position of the deflected protons, the path-integrated electromagnetic fields can be reconstructed. 
To simplify the deflection measurement, a metallic mesh that splits the proton beam into beamlets can be inserted between the proton source and plasma of interest \cite{Nilson:2006,Li:2006,Li:2007b,Petrasso:2009, Willingale:2010,Rosenberg:2015}.
Mesh-based proton deflectometry has been used extensively to study electromagnetic fields using two types of proton sources. Exploding pusher fusion monoenergetic protons \cite{Seguin:2003} have been used to study Biermann battery fields and magnetic reconnection \cite{Petrasso:2009,Li:2007b,Rosenberg:2015}, while protons generated through Target Normal Sheath Acceleration (TNSA) \cite{Wilks:2001} have been used to study self-generated magnetic fields in targets \cite{Willingale:2010} and $\sim$kT magnetic fields in laser-driven capacitor coils \cite{Santos:2015,Bailly:2018}. 

The accuracy of the magnetic field measurement using this technique strongly depends on knowing the undeflected beamlet positions. In previous experiments that was deduced from regions of the mesh unaffected by magnetic fields or obtained in a separate null shot (i.e. zero magnetic field and no plasma) measurements. However, in many experiments the magnetic fields occupy a volume larger than the measurement field of view, precluding measurements of the unperturbed mesh. Alternatively, a separate measurement with no magnetic field implies additional errors and requires precise alignment and reproducibility of the mesh between shots

Here we report the design and implementation of a platform at the Omega laser facility for measuring non-uniform magnetic fields using proton deflectometry that includes an \textit{in situ} X-ray reference of the mesh \cite{johnson:2021}. The technique is based on using X-rays and protons generated from the same source. This allows simultaneous measurements of the proton deflections and a reference image of the unperturbed mesh using X-rays, which is crucial for accurately calculating the electromagnetic fields.  In this paper, we discuss experiments using two different proton/X-ray sources -- imploded D$^3$He fusion sources and laser-driven TNSA sources -- and demonstrate the successful execution of both.  

Using Monte-Carlo simulations, we also show how the technique can be optimized further by varying the mesh parameters to obtain higher quality X-ray and proton images. The simulations show that the enhanced contrast of proton and x-ray images can be achieved by using large gold mesh. Based on simulations, we also provide a simple guide for obtaining good quality images for different mesh sizes.

The paper is organized as follows. First in Section 2.1 we describe the experimental setup and results obtained with exploding pusher targets at the OMEGA laser facility and discuss the accuracy of the magnetic field measurements using this technique. Section 2.2 describes the implementation of the technique using a TNSA proton source at the OMEGA EP laser facility. Section 3 focuses on the modeling the results and optimizing the technique of the technique using FLUKA Monte-Carlo simulations for both OMEGA and OMEGA EP experiments. In addition, we present analytical calculations for enhancing X-ray contrast in Section 3.3.

\section{Experimental results}

In this section we describe our technique that pairs proton deflectometry with a simultaneous X-ray reference as implemented at the OMEGA and OMEGA EP laser facilities, using two different types of proton source: exploding pusher targets that produce mono-energetic protons at 3 and 15 MeV via D-D and D-$^3$He fusion reactions reactions\cite{Seguin:2003}, and broad-spectrum protons produced through the laser-driven target-normal-sheath acceleration (TNSA) process\cite{Wilks:2001}. 
Exploding pusher protons have an isotropic distribution, relatively homogeneous flux \cite{Seguin:2003,TJohnson:2021}, a typical time duration of $\sim$ 100 ps, and a spatial resolution of order 50 $\mu$m. The detector typically consist of two layers of CR39 films for 3 and 15 MeV protons respectively. TNSA protons are produced in a cone with a half angle of $\sim$ 30$^\circ$ and have relatively inhomogeneous flux \cite{Snavely:2000}. The detector in this case is consist of several RCF layers and metallic filtering which allows to detect desirable energy of protons up to cut-off energy (i.e. 5,6,10,20)  and achieve temporal resolution of 5-10 ps and spatial resolution of tens of $\mu$m.

\subsection{Implementation for D$^3$He backlighter protons at OMEGA}

\begin{figure}[h!]
\centering\includegraphics[width=0.95\textwidth]{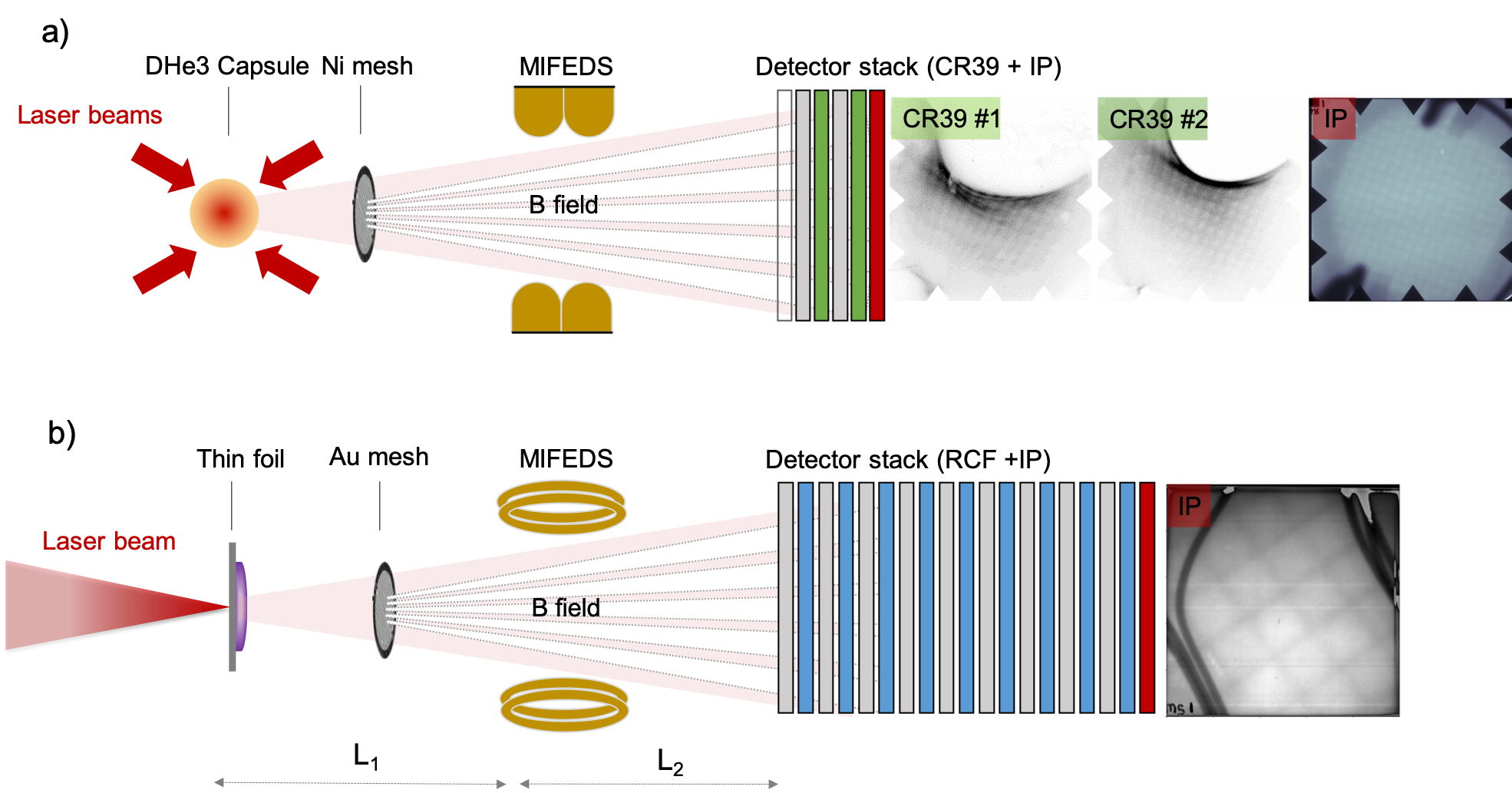}
\caption{ Schematic of the experimental setup with example raw image data (not to scale). \textbf{a)} At OMEGA laser facility. The detector stack was composed of CR-39 (Columbia Resin 39) for 3 MeV and 15 MeV protons (green color) and an image plate (IP) (red color) in the back of the stack for an X-ray image of the Ni mesh.  \textbf{b)} At OMEGA EP laser facility . The detector stack was composed of RCF (Radiochromic film) layers (blue color) detecting different energies of protons and an IP in the back of the stack.}
\label{setup}
\end{figure}

A schematic of the experimental setup at the OMEGA laser facility is shown in Fig.~\ref{setup} a). A quasi-static magnetic field was produced by the Magneto-Inertial Fusion Electrical Discharge System (MIFEDS) \cite{Fiksel:2015}. A MIFEDS current of 22 kA was used to generate a magnetic field strength of 4 T at the midpoint between the coils over a volume of $\approx$ 5 mm$^3$. The MIFEDS design was modeled using COMSOL software based on Biot-Savart calculations to obtain initial magnetic field for the region between the coils for comparison with experimental data. The initial magnetic field was spatially non-uniform due to the design of the coils, with the field strength increasing from the midpoint toward the coils.

Proton radiography was employed to measure the magnetic field in the region between the coils. Imploded D$^3$He-filled capsules produced an isotropic point-like (20-50 $\mu$m radius) source of 3 MeV and 14.7 MeV protons as fusion products, as well as continuum X-rays \cite{Seguin:2003}. 
The capsule was 420 $\mu$m in diameter and filled with a
mixture of D$_2$ and $^3$He gas and was imploded with 19 OMEGA drive beams, each delivering an energy of 500 J.
The capsule was positioned $L_1 =$~10~mm from the target-chamber-center (TCC), along the proton radiography axis, and the detector stack was positioned a distance $L_2 =$~154~mm on the opposite side of TCC.
The protons were split into beamlets using a Ni mesh positioned 4 mm from the backlighter source. The mesh featured 125 $\mu$m pitch, 90 $\mu$m mesh opening and a bar thickness of 35 $\mu$m.  
The detector stack used in the experiment (Fig. \ref{setup} a ) consisted of two CR-39 detectors, one each for the 3~MeV and 14.7~MeV protons, and an image plate detector (IP, Fujifilm SR-type) sensitive to X-rays. In order to match the position of the Bragg peak with the CR-39 detectors, a 15~$\mu$m Ta filter was placed in front of the first CR-39, and a 200 $\mu$m Al filter was placed in front of the second CR-39. For this stack the minimum energy of X-rays reaching the image plate was estimated to be $h\nu \gtrsim~25$~keV ($1/e$ attenuation) using X-ray transmission calculations and Monte-Carlo simulations. A jagged fiducial frame in the front of the stack, which left a "tooth" pattern around the border, was introduced to facilitate the alignment of the CR-39 and IP images during data processing. The CR-39 were processed and scanned under a microscope by established techniques \cite{Seguin:2003}, producing a map of proton counts per pixel, while the IP was scanned with a commercial Fujifilm Typhoon FLA-7000 scanner with a resolution of $r = 25$ $\mu$m and sensitivity of $S = 1000$.

\begin{figure}[h!]
\centering\includegraphics[width=0.9\textwidth]{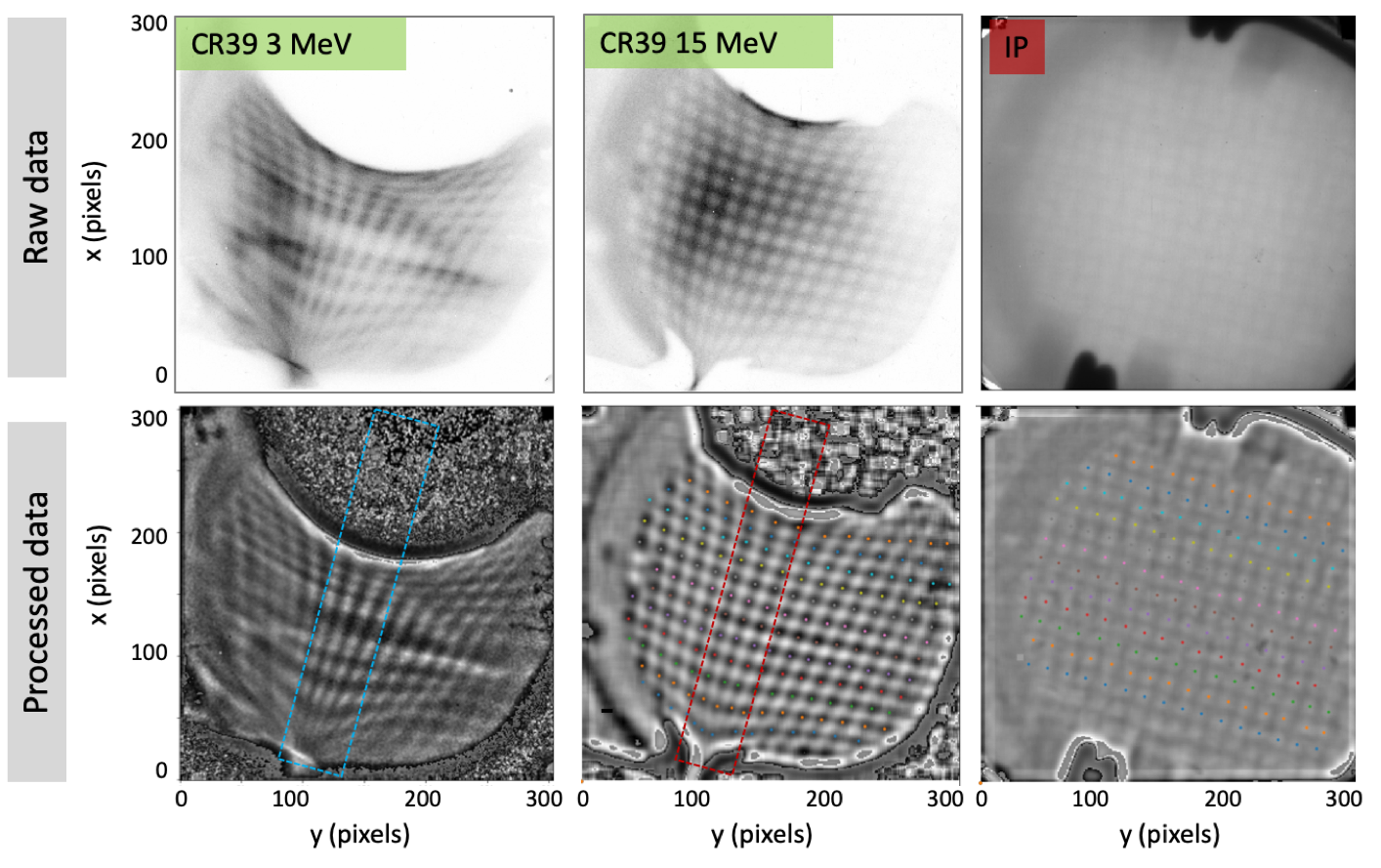}
\caption{(Top) Raw X-ray image of the mesh and 3 and 15 MeV proton images obtained for vacuum magnetic field (no plasma). (Bottom) Corresponding processed images with identified beamlet positions using a semi-automatic routine. The three columns of beamlets in the mid-plane (between the MIFEDS coils) indicated by the blue (3 MeV) and the red (15 MeV) dashed boxes are used to calculate path-integrated magnetic field profiles in Fig.\ref{bdl}.}
\label{rawdata}
\end{figure}

The vacuum magnetic field was measured using both a conventional technique, in which deflected and undeflected proton beamlets are measured on separate shots (one being a reference shot with no magnetic field), and the new technique in which proton and X-ray images provide both deflected and undeflected beamlet positions on a single shot. 
The proton radiographs obtained using the new technique are shown in the Fig. \ref{rawdata}
The path-integrated magnetic field was obtained using the standard theory of proton deflectometry \cite{Kugland:2012}.

Protons propagating through the magnetized volume are deflected by the Lorentz force with a deflection angle given by:
\begin{equation}\label{eq:alpha}
\Delta \vec{\alpha}=\frac{e}{m_p v_p^2}\int{(\vec{E} +\vec{v}\times\vec{B}) dl},
\end{equation}
where $v_p = \sqrt{2 E_p / m_p}$ is the proton velocity corresponding to an energy $E_p$.
In these experiments, there is no significant contribution of electric field to the deflection of the protons since the electric field strength is negligible. 
The deflection angle can be directly calculated  from the experimental measurements by:

\begin{equation}\label{eq:final_Bdl}
 \Delta \vec{\alpha} =  \frac{L_1+L_2}{L_1 L_2} (\vec{d_2} - \vec{d_1}).
\end{equation}
where $d_{2}$ is the deflected beamlet position obtained from the proton image and $d_{1}$ is an undeflected beamlet position obtained from the X-ray image. Here both d1 and d2 are given in terms of the coordinate system of the object plane, which is related to the observation on the detector by the magnification $M = (L_1 + L_2) / L_1$ between the object and detector planes, since we use a magnified point-source geometry. 
Then the path integrated magnetic field can be calculated by:
\begin{equation}\label{eq:Bdl}
\int{d\vec{l}\times\vec{B}} = \frac{m_p v_p}{e} \frac{L_1+L_2}{L_1 L_2} (\vec{d_{2}} - \vec{d_1}).
\end{equation}

The raw experimental data for the vacuum magnetic field is shown in Fig. \ref{rawdata} (top) and features the 3 MeV proton image (left), the 15 MeV proton image (center), and the X-ray image of the mesh (right).
A detailed explanation of the data analysis of this diagnostic technique is presented in \cite{johnson:2021}, and we summarize the main components here.

The analysis of data was based on the following steps: the raw images were first processed to enhance their contrast; the proton images were then overlapped with the X-ray image of the mesh; and then the deflected (CR-39) and undeflected (IP) beamlets were located by a semi-automatic routine. 

The post-processed images with enhanced contrast are shown in Fig. \ref{rawdata} (bottom). As a measure of the image quality, we introduce an experimental beamlet signal to noise ratio defined as the ratio between the minimum (maximum) range value of the beamlets to the rms value of the sub-beamlet-scale noise. 

The original contrast of the 3 MeV proton image was estimated as  $\sim$ 7, the 15 MeV proton image as $\sim$ 8 , and the X-ray image as $\sim$ 2. After post processing the contrast was enhanced to $\sim$ 42 , $\sim$ 60 and $\sim$ 19, respectively.

\begin{figure}[h!]
\centering\includegraphics[width=0.8\textwidth]{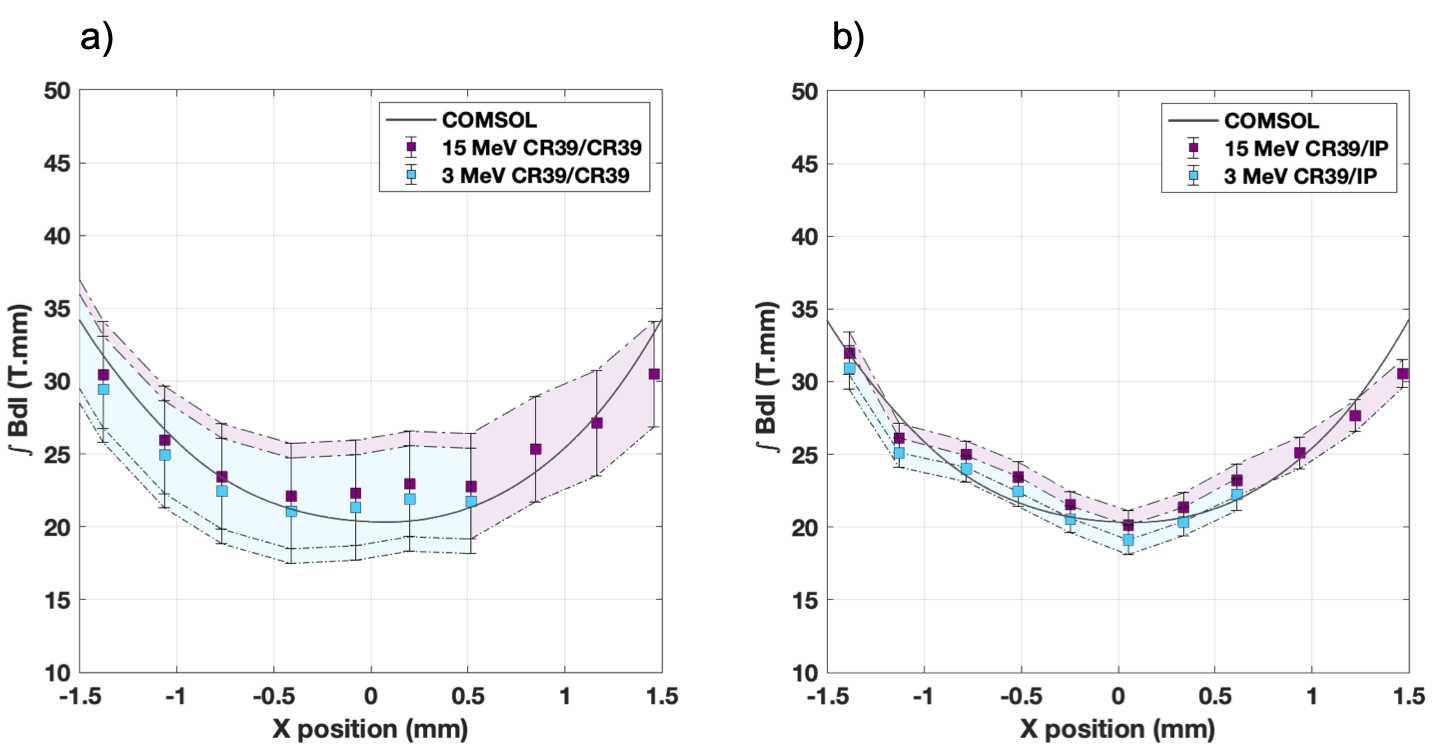}
\caption{The averaged path-integrated magnetic field obtained using the three columns of beamlets in the mid-plane as shown in Fig. \ref{rawdata} (bottom) using \textbf{a)} the conventional CR-39/CR-39 technique and \textbf{b)} our CR-39/IP technique.  Both are compared with COMSOL simulations. }
\label{bdl}
\end{figure}

The averaged path-integrated magnetic field calculated using the 3 and 15 MeV proton images is shown in Fig. \ref{eq:Bdl}. Three columns (see Fig. \ref{rawdata} (bottom)) were averaged together to obtain the path-integrated magnetic field in the region between the coils. The results were compared with COMSOL simulations of the magnetic field based on the known coil geometry. The total error of the measurement was calculated as $\sigma_{tot}$ = $\sqrt{\sigma^2 + \sigma_{sys}^2}$, where $\sigma$ is the statistical error obtained from averaging lineouts, and $\sigma_{sys}$ is a systematic error. 

As mentioned above, this new technique was evaluated against the traditional  method of comparing a proton image from two separate shots (CR-39/CR-39). Fig. \ref{bdl} shows the comparison of the results a) using the CR-39/CR-39 technique and b) using the CR-39/IP technique.  As can be seen, the accuracy of the magnetic field measurement improves by a factor of $\sim$4. The CR-39/IP technique has an average error of $\sim$1.2 T~mm, while the CR-39/CR-39 technique has an error of $\sim$5.7 T~mm.
This higher error for the CR-39/CR-39 technique arises from several sources, including the error from aligning the two CR-39 detectors, the error of the beamlet positioning, and the relative alignment error in the mesh position between two separate shots (estimated to be 25 $\mu$m in the mesh plane). While the CR-39/IP technique has errors from the aligning of the CR-39 and IP and from locating the beamlets, there is no error from the positioning of the mesh, since the measurement is performed in one shot using the same mesh.
The technique can still be improved by reducing  the error on the alignment procedure of the CR-39 and IP images, and by reducing the error on the beamlet location.
Two main sources of uncertainty in locating the center of the beamlets are the mesh contrast and the image signal-to-noise ratio. While it is quite difficult to improve the signal-to-noise ratio experimentally, the mesh contrast can be increased,  which would facilitate identification of the center position.

\subsection{Implementation for TNSA protons at OMEGA EP}

The performance of the CR-39/IP technique was also demonstrated using laser-driven TNSA protons at the OMEGA EP laser facility. A schematic of the experimental setup is shown in Fig. \ref{setup} b), which can be compared to the OMEGA experimental setup in Fig. \ref{setup} a). The main differences between the two experimental setups are the proton source characteristics and proton detector stack. While the backlighter implosion at OMEGA produces 3 and 15 MeV mono-energetic protons that are detected by two CR-39 detectors, the laser-driven TNSA protons have a broadband spectrum up to 40 MeV and are detected by a radiochromic film (RCF) stack designed to detect certain proton energies in each film layer. From another side, x-rays are produced through Bremsstrahlung emission from laser-driven hot electrons propagating through the thin foil target.  The detector stack is much thicker with a total thickness of  $\approx$ 6 mm, which significantly attenuates the X-rays  that that reach the IP positioned at the back of the stack.

In the OMEGA EP experiment a high-intensity laser pulse ($\sim$ 1 ps, 50 J) was focused onto a 20 $\mu$m thick Au foil with a focal spot of 50 $\mu$m to produce a TNSA-like proton spectrum with a cut-off energy of 25-30 MeV. The energy of EP backlighter was limited to 50 J due to the debris shield of the focusing off-axis parabola that has to be installed for the experiments with MIFEDS coils. The MIFEDS coils were designed using COMSOL software to achieve a magnetic field of $\sim$ 10 T in the mid-plane between the coils.

The proton radiography setup was similar to the OMEGA experiment and utilized the same distances $L_{1}$ and $L_{2}$. An Au mesh was positioned 4 mm away from proton source. The mesh featured 340 $\mu$m pitch, 285 $\mu$m mesh opening and a bar thickness of 55 $\mu$m. The mesh type, detector stack composition, and IP position in the stack were designed using FLUKA Monte-Carlo simulations described in the next section.
The detector stack consisted of 12 layers of HD-V2 RCF film, 12 layers of Al filters of various thickness placed between each film layer, and a MS-type IP placed in the back of the stack. The RCF stack used in the experiment was designed to detect proton energies between 5.5 and 32.5 MeV. The total thickness of Al filters in the stack was 4.2 mm. 
The IP plate was placed directly after the last film layer with no additional filtering. Fig. \ref{setup} b) shows the raw experimental X-ray image of the mesh obtained on the IP. The contrast ratio of the X-ray image is $\approx$ 5.
The proton spectrum on this shot was observed to cut off by the 6th stack, corresponding to energy 15 MeV.  Therefore, the image on the IP was an x-ray image, which is obvious due to its regular mesh structure.  This illustrates a successful implementation of this CR39/IP technique for TNSA protons.

\section{Modeling and optimization of the technique}

In this section we use the FLUKA simulation code \cite{Bohlen:2014,Ferrari:2005} to perform a numerical study of the performance of the CR-39/IP technique using exploding pusher (Sec 2.1) and laser-driven TNSA (Sec 2.2.) protons. FLUKA is a  multi-purpose Monte-Carlo code for simulating the transport and interaction of charged particles and photons over a wide range of energies (from keV to cosmic rays) with materials. It also includes particle scattering and transport in electromagnetic fields.

\subsection{Modeling of OMEGA experiment}

The simulations were performed using several different mesh sizes and materials in order to quantitatively compare the resulting image contrast and blurring, with the goal of determining the optimum mesh parameters for the measurement.

\begin{figure}[h!]
\centering\includegraphics[width=1\textwidth]{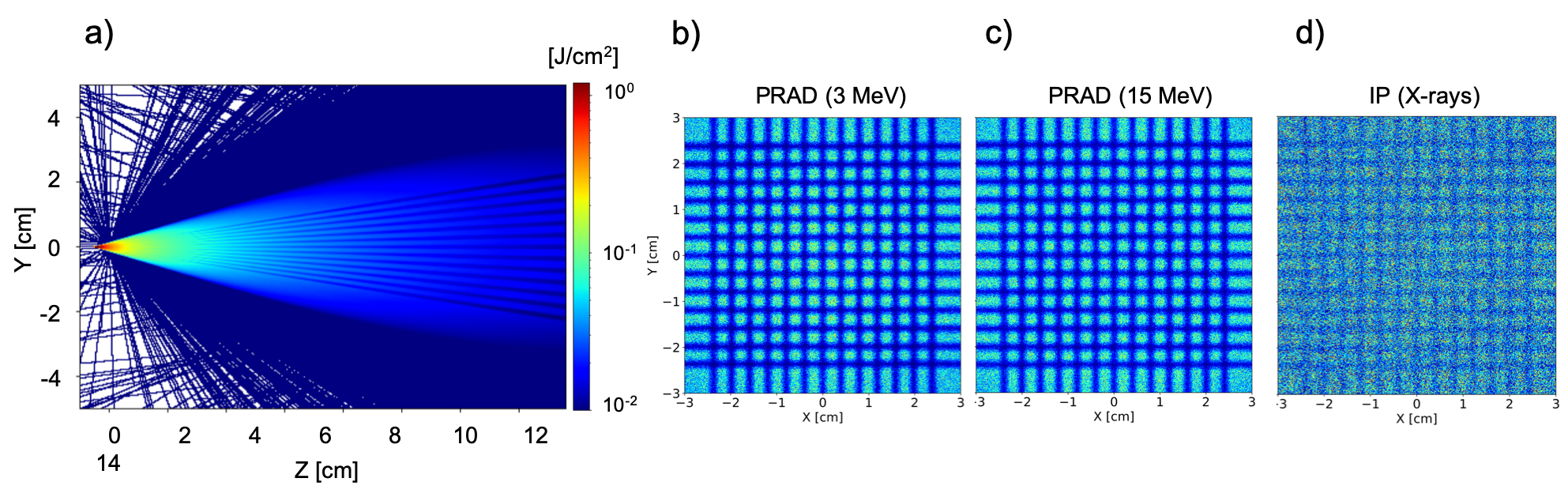}
\caption{3D FLUKA simulations of proton and X-ray radiography of a Ni mesh modeled after the OMEGA experimental setup. \textbf{a)} Proton flux along the propagation direction from the source to detector at $x = 0$. \textbf{b)} Synthetic proton radiography of the mesh obtained with 3 MeV protons. The data was extracted at the position of the first CR-39 detector ($z = 15.065$ cm) after passing through a 15 $\mu$m Ta filter. \textbf{c} Synthetic proton radiography of the mesh obtained with 15 MeV protons. The data was extracted at the position of second CR-39 detector ($z = 15.58$ cm) after protons passed through a 15 $\mu$m Ta filter, the CR-39 for 3 MeV protons, and 200 $\mu$m Al filter. \textbf{d)} Synthetic X-ray image of the mesh obtained by X-ray spectrum. The data is extracted at the position of the IP at the back of the detector stack ($z = 16.2$ cm)}
\label{fluka_omega}
\end{figure}

The simulations were performed using the experimental distances for proton radiography, $L_{1}$ = 10 mm and $L_{2}$ = 154 mm, mesh parameters (125 $\mu$m mesh period), and a proton source defined as follows: 15 or 3 MeV protons with source size of $r$ = 50 $\mu$m, initial divergence of 550 mrad, and 5 $\times$ 10$^6$ primary particles located 4 mm away from mesh. While in the experiment protons have an isotropic distribution, we limit the divergence of the source to reduce computational time. As an X-ray source we used an exponential spectrum $I_{0}(E) = \exp^{-E/kT}$, where $T = 5$ keV is the characteristic temperature, with the same spatial and angular parameters as for the proton source.
The simulations replicated the experimental detectors, with a CR-39 and IP stack, where a Ta filter is placed in front of the first CR-39, an Al filter is placed between first and second CR-39, and the IP is at the back. CR-39 and IP material and chemical compositions were specified as provided by the manufacturer. Fig. \ref{fluka_omega} shows the results of the simulations featuring (a) the proton flux along the propagation axis $z$, (b) the 3 MeV proton flux at the detector plane of the first CR-39, (c) the 15 MeV proton flux at the second CR-39, and (d) the X-ray flux at the IP.

\begin{table}
\begin{center}
\begin{tabular}{lccc}
\hline
\textbf{Mesh type} & \textbf{Large} & \textbf{ Medium}  & \textbf{Small}  \\
\hline
Mesh period &  340 $\mu$m & 125 $\mu$m & 83 $\mu$m \\
Hole &  285 $\mu$m &90 $\mu$m & 58 $\mu$m\\
Bar width &  55 $\mu$m &35 $\mu$m & 25 $\mu$m\\
Mesh period at object plane  & 848 $\mu$m  &312 $\mu$m &207 $\mu$m\\
\hline
\end{tabular}
\caption{Relevant dimensions of the meshes used in the simulations.}
\label{table:parameters}
\end{center}
\end{table}

As a measure of the quality of the CR-39/IP technique, we introduce two parameters: the contrast ratio and the degree of blurring. The contrast ratio is defined as the ratio between the particle fluence through the mesh hole (maximum signal) versus through the mesh bar (minimum signal), calculated at each mesh period of the synthetic image to obtain an average ratio. The degree of blurring is a measure of how much a mesh bar edge is blurred into a mesh hole, since this additional noise can reduce the precision of identifying the center of the mesh hole.
The blurring effect is defined as $b =  (\Delta Y - \Delta X )/ \Delta X \times 100 \%$, where $\delta Y$ is the size of the mesh obtained from the synthetic image and $\delta X$ is the true size of the mesh on the detector as calculated from simulations using a point-like source. 
The simulation results presented in Fig. \ref{fluka_omega} b-d) show a contrast ratio of $\sim$ 4 for the 3 MeV proton image, $\sim$ 3.5 for the 15 MeV proton image, and $\sim$ 1.3 for the X-ray image, respectively, and blurring of $\sim$ 24 $\%$. While the contrast ratios are quite similar for the 3 and 15 MeV proton images, the X-ray image has a reduced contrast by a factor of 2 - 2.5 that qualitatively agrees with the experimental observations at OMEGA.

\begin{table}
\begin{center}
\begin{tabular}{lccc}
\hline
\textbf{Parameter} & \textbf{3 MeV p }& \textbf{15 MeV p}  & \textbf{X-ray} \\
\hline
\multicolumn{4}{c}{Large Mesh }\\
\hline
$\delta Bdl$ of one mesh unit        & 22.9 T~mm & 53 T mm & None \\
Contrast ratio Ni & 15.5 &14 &1.2  \\
Contrast ratio Au & 17 &16.1 &3.6  \\
Blurring & 13 $\%$ &14 $\%$ & 13 $\%$ \\
\hline
\multicolumn{4}{c}{Medium Mesh }\\
\hline
$\delta Bdl$ of one mesh unit        & 8.4 T~mm & 19.5 T mm & None  \\
Contrast ratio Ni & 4  &3.5 &1.28  \\
Contrast ratio Au & 5 &4.8 &2.25  \\
Blurring & 24 $\%$ & 25.5 $\%$ & 24 $\%$ \\
\hline
\multicolumn{4}{c}{Small Mesh }\\
\hline
$\delta Bdl$ of one mesh unit        & 5.5 T~mm & 12.9 T mm & None \\
Contrast ratio Ni & 2.3  &1.57  & 1.05 \\
Contrast ratio Au & 2.46 & 2.3 & 1.8 \\
Blurring & 38 $\%$ & 40 $\%$ & 38 $\%$ \\
\hline
\end{tabular}
\caption{Parameters of the performance of the CR-39/IP technique calculated for different mesh sizes and materials.}
\label{table:parameters}
\end{center}
\end{table}

\begin{figure}[h!]
\centering\includegraphics[width=0.7\textwidth]{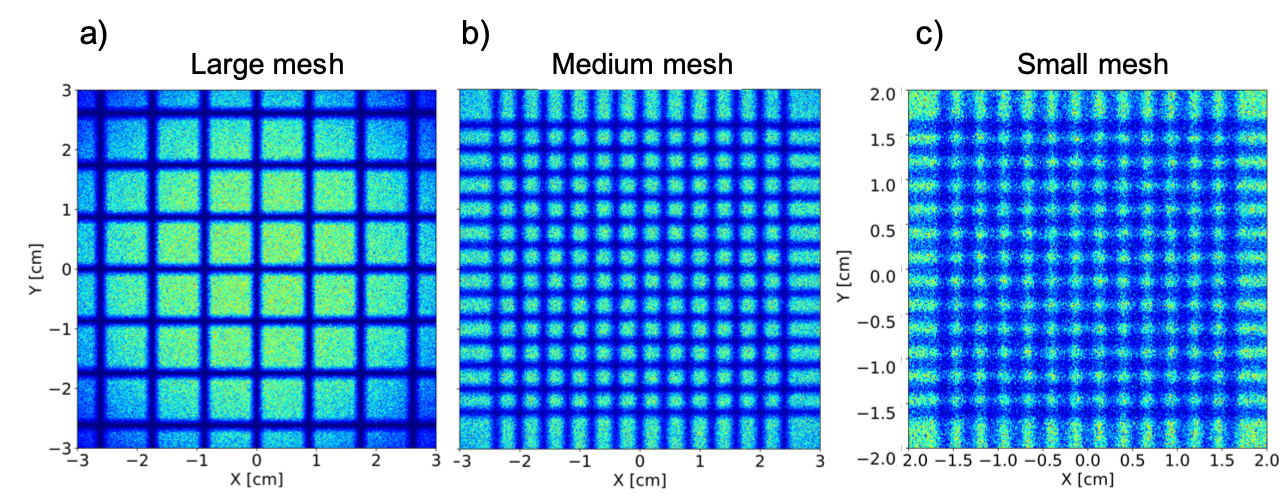}
\caption{Comparison of 15 MeV proton images of the Au mesh with the different pitch sizes shown in Table 1. \textbf{a)} Large Mesh. \textbf{b)} Medium mesh (used in the experiment). \textbf{c)} Small mesh.}
\label{fluka_mesh}
\end{figure}

In order to find optimal mesh parameters that maximize contrast and minimize blurring for both the proton and X-ray images, we performed a set of simulations modeling a variety of commercially-available meshes, including small, medium, and large mesh periods and Ni and Au mesh material, keeping the main parameters of the proton radiography setup such as source size, geometrical distance, and detector stack the same.
Table 1 summarizes the relevant dimensions of the large, medium and small meshes used in the simulations. The mesh period in the object plane was calculated using a geometrical magnification factor of 2.5.

While using different sizes of the mesh, it is also important to note a parameter that defines a quantitative measure of the beamlet deflection by the magnetic field: the change in path-integrated field $\delta Bdl$ corresponding to one mesh unit. The choice of the mesh size also depends on the strength of the magnetic field in experiment that needs to be measured. As the mesh unit sizes increase, the corresponding calculated path-integrated magnetic field will also increase.

Table 2 shows a comparison of the contrast ratio, blurring, and $\delta Bdl$ of one mesh unit for the CR-39/IP simulations using large, medium and small meshes.
Fig. \ref{fluka_mesh} shows the synthetic proton images obtained with 15 MeV protons with a source size of 50 $\mu$m using large, medium and small meshes, which clearly indicate increased blurring when using smaller mesh size. 
The blurring effect, as expected, mostly depends on the size of the mesh with respect to the proton source size and has a lower sensitivity to scattering that depends on the incident proton energy. 

The scattering angle of protons with given energy $E$ in material with areal density $A$ can be estimated as  $\zeta$ $\approx$ $\sqrt{A [\text{g/cm}^{3}]}/ E[\text{MeV}]$ \cite{Volpe:2011,Rossi:1941}. For a Ni mesh, 15 MeV protons have a scattering  angle of $\approx$ 0.5 $^\circ$, while 3 MeV protons have an angle of $\approx$ 2.8 $^\circ$. For an Au mesh, the angles are $\approx$ 0.98 $^\circ$ and $\approx$ 4.8 $^\circ$ for 15 MeV and 3 MeV protons, respectively. The increased scattering angle for 3 MeV protons results in increased blurring of the synthetic proton image by on average 1-2 $\%$. While the image gets more blurry, it simultaneously loses contrast between the mesh bar and mesh hole. 
The contrast ratio of the synthetic proton image can be enhanced by a factor $\sim$4 by using a larger mesh compared to the medium Ni mesh used in the OMEGA experiment described in Section 2.1.
Additionally, the quality of the X-ray image can be enhanced by a factor of 2-3 using a Au mesh, since more X-rays are absorbed in Au compared to Ni. 
These enhancements significantly reduce the uncertainty in beamlet location, thereby improving the accuracy of the magnetic field measurement by approximately a factor of 2 over the results presented in Fig. \ref{bdl} b).

\subsection{Modeling of OMEGA EP experiment}

FLUKA simulations were also performed for the design of the detector stack for the OMEGA EP experiment described in Section 2.2. The main goals of the simulation were to identify the optimum position of the image plate in the RCF stack and to define the optimum mesh material and size in order to obtain the maximum contrast of the X-ray image. 
The simulations were performed with an X-ray source with a central energy of 35 keV and a bandwidth of 20 keV (FWHM), a source size of $r$ = 50 $\mu$m, and an initial divergence of 900 mrad  with 1 $\times$ 10$^7$ primary particles.  The source was located 0.4 cm away from the mesh. 
The simulation setup replicated the experimental distances $L_{1}$ and $L_{2}$, as well as the RCF stack that consisted of 12 RCF layers, 12 Al filters of various thickness, and an IP in the back of the stack. The RCF material and chemical composition were specified by the manufacturer.

\begin{figure}[h!]
\centering\includegraphics[width=0.8\textwidth]{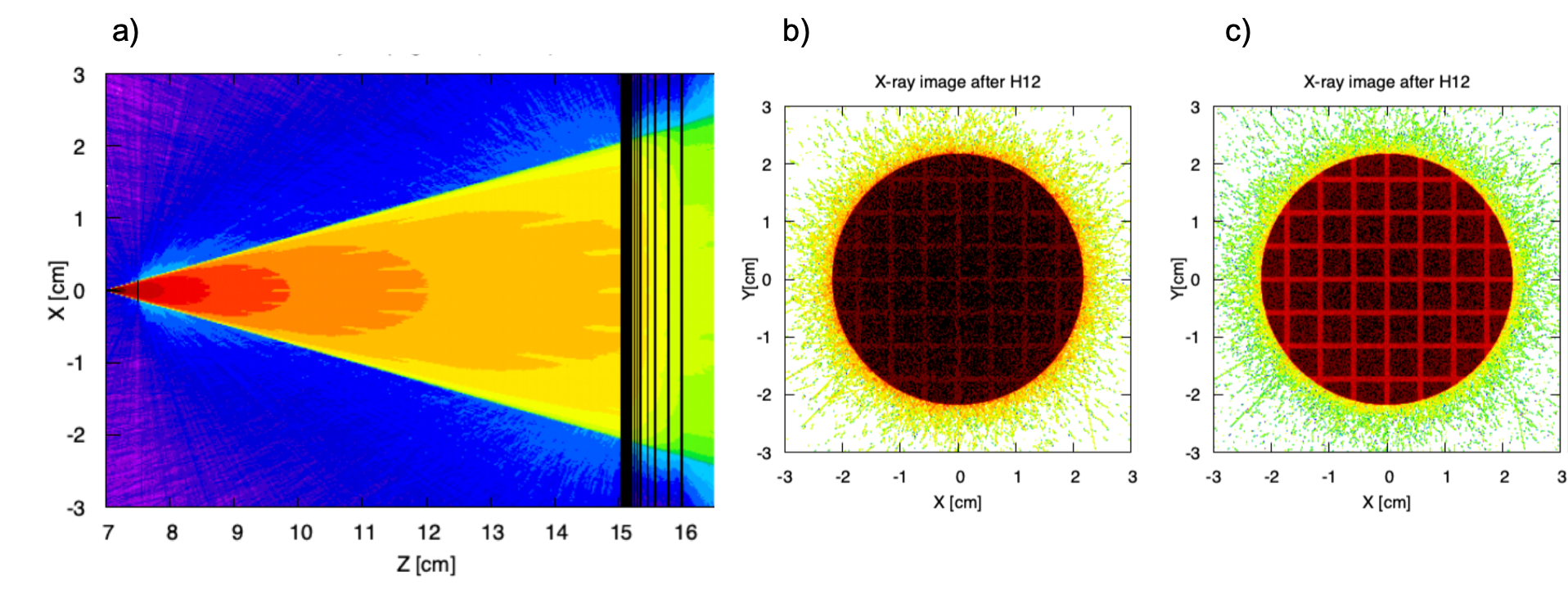}
\caption{FLUKA Monte-Carlo simulations of X-ray radiography modeled after the OMEGA EP experiment. \textbf{a)} X-ray flux along the propagation direction $z$ at $x = 0$. X-ray flux was extracted at the position of the IP ($z = 16$ cm) for a large \textbf{b)} Ni mesh  and \textbf{c)} Au mesh. }
\label{fluka_ep}
\end{figure}

Two different sets of meshes were compared in order to determine the best contrast: large Ni and Au meshes and medium Ni and Au meshes as described in Table 1.
Fig. \ref{fluka_ep} a) shows the X-ray flux along the propagation axis with the geometry of the RCF stack overlaid, as well as the X-ray flux extracted at the position of the IP for b) the Ni mesh and c) the Au mesh. An optimal placement of the IP behind the layers of RCF was determined by a trade-off between the maximum recorded proton energy and the contrast ratio of the X-ray image on the IP. 
As a result of the simulations, we can observe an increase of contrast by a factor of 6 when using the Au mesh compared to the Ni mesh.
The pre-shot simulations also qualitatively agree with the experimental results at OMEGA EP.
 
\subsection{Analytical modelling}

\begin{figure}[h!]
\centering\includegraphics[width=0.45\textwidth]{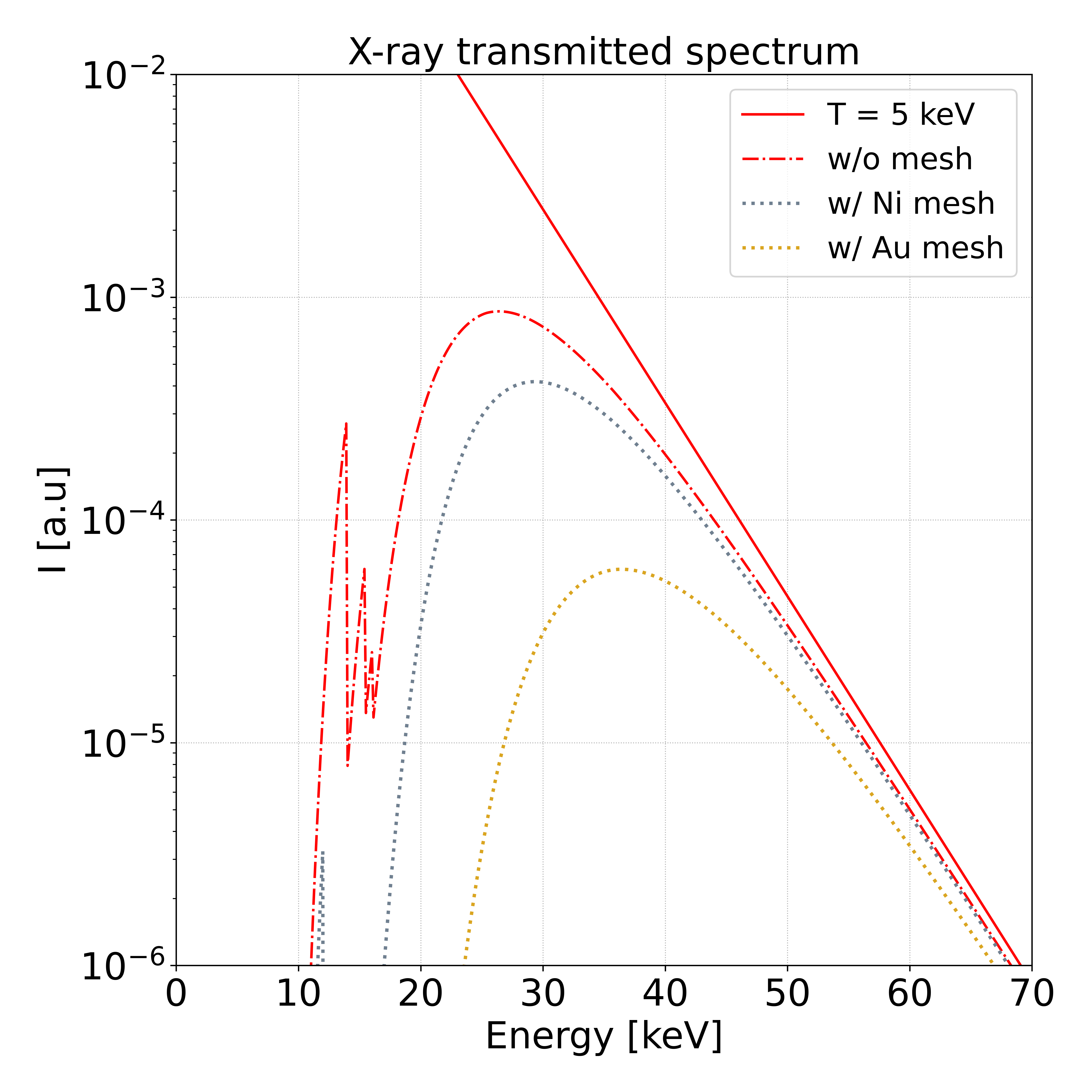}
\caption{Comparison of calculated X-ray spectra, including the initial spectrum with characteristic temperature of $T = 5$ keV (solid red line), the spectrum transmitted through the CR-39 detector stack (dashed red line), the spectrum transmitted through the detector stack and Ni mesh (grey dashed line), and the spectrum transmitted through the Au mesh (gold dashed line).}
\label{xray_transmission}
\end{figure}

To investigate additional methods for further improving the X-ray image contrast, an analytic modeling study was performed. The detected X-ray spectrum was calculated for the OMEGA experiment setup using the CR-39 stack and Au and Ni mesh materials.
The initial X-ray source spectrum generated in the implosion was represented by an exponential spectrum, $I_{0} = \exp(-E/kT)$, where T is a characteristic temperature observed as $T = 2 - 5$ keV for OMEGA implosions \cite{Adrian:2021}. For further calculations we take $T = 5$ keV.
The detected spectrum of the X-rays passing through the mesh can then be written as:
\begin{equation}
I_{det} = I_{0}(E)\times T_{mesh}(E) \times T_{stack}(E)\times G(E),
\end{equation}
where $T_{mesh}(E) $ is the X-ray transmission through the mesh bar, $T_{stack}(E)$ is the X-ray transmission through the CR-39 stack composition, and $G(E)$ is the IP response function for different types of IPs (MS, SR, TR) taken from \cite{Meadowcroft:2008}.
The X-ray transmission for energies of 0 - 100 keV is calculated using tables from the Center for X-ray Optics \cite{lbl:0000}.
Fig. \ref{xray_transmission} shows the initial X-ray spectrum and the X-ray spectra transmitted through just the CR-39 detector stack (i.e. through a mesh hole), the Ni mesh and detector stack (i.e. through a mesh bar), and the Au mesh and detector stack. 

\begin{figure}[h!]
\centering\includegraphics[width=0.9\textwidth]{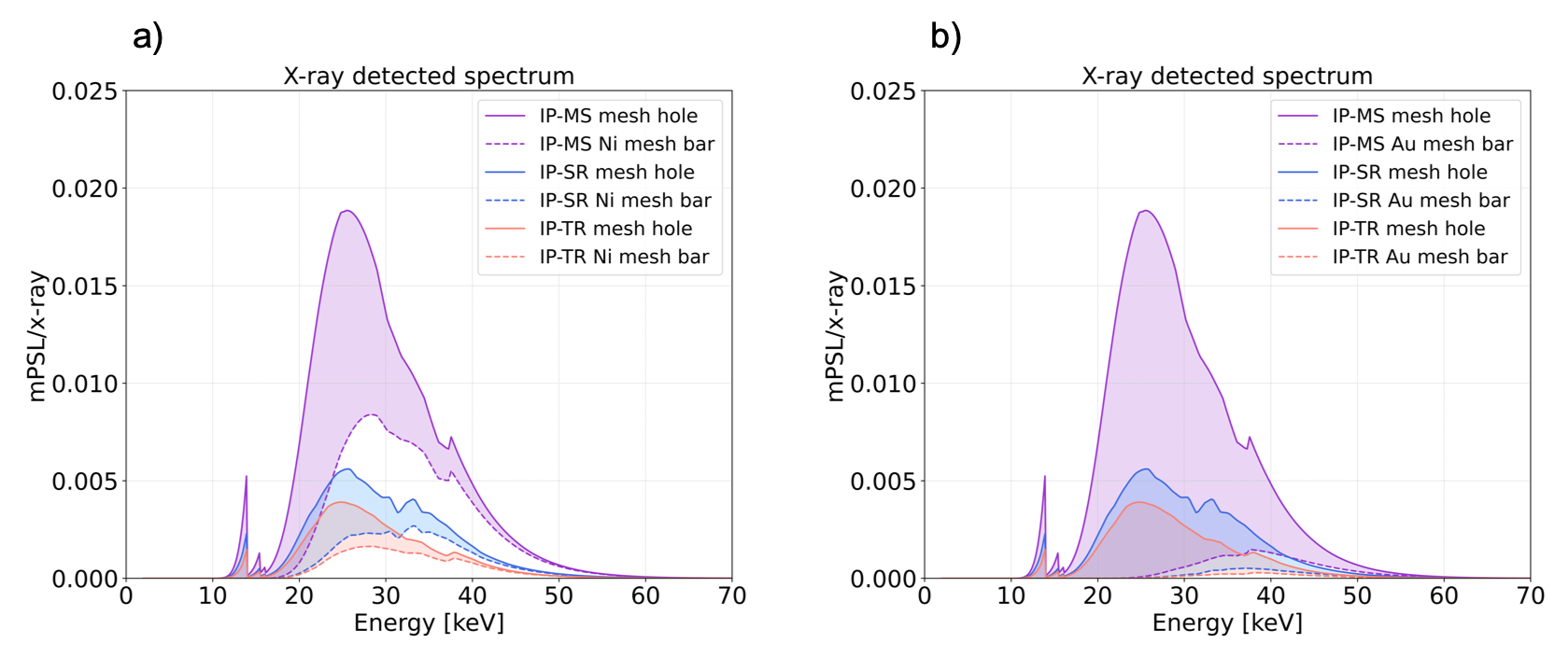}
\caption{Calculated X-ray spectrum detected for different IP types (MS,SR,TR) obtained using the IP response functions from \cite{Meadowcroft:2008}. \textbf{a)} Comparison for a Ni mesh, including transmission through the Ni mesh and detector stack (dashed line) and transmission through the detector stack only (solid line) for IP-MS (purple), IP-SR (blue), IP-TR (red). \textbf{b)} Similar comparison for a Au mesh (dashed line). The shaded area indicates the contrast of the mesh represented as a difference between the transmitted spectrum through the detector stack with Ni or Au mesh and the reference spectrum transmitted through the detector stack only.}
\label{xray_detected}
\end{figure}

The detected spectrum through the mesh hole and Au and Ni meshes can then be calculated by multiplying the transmission spectrum by the IP response function for different IP types. The Fig. \ref{xray_detected} shows the calculated detected spectrum on the IP-MS , IP-SR, and IP-TR for X-rays passing through the mesh hole and through the mesh bar. 
The ratio of the detected spectrum through the mesh hole and through mesh bar, indicated by the shaded area in the plot, can be called as contrast ratio and is defined as
\begin{equation}
C = \frac{\int I_{mesh-hole}(E)dE}{\int I_{mesh-bar}(E)dE}.
\end{equation}
As one can see, there is a clear enhancement of contrast using an Au mesh. The contrast ratio calculated for a Ni mesh (Fig. \ref{xray_detected} a) is 2.04 for IP-MS, 2 for IP-SR, and 2.12 for IP-TR, while for a Au mesh (Fig. \ref{xray_detected} b) it is 12.69, 11.69, 13.63, respectively, indicating an increase of contrast by a factor of 6 using a Au mesh. Note that the use of the IP-TR and Au mesh can also enhance contrast by 20 $\%$.  There is no major difference observed using different IP types for the case of the Ni mesh.

It is worth noting that the resulting contrast indeed depends on the initial X-ray spectrum and its characteristic temperature. While the typical temperature is usually estimated as 5 keV for OMEGA implosions, it can also vary within a several keV range for laser-driven X-ray sources. 

\begin{figure}[h!]
\centering\includegraphics[width=0.9\textwidth]{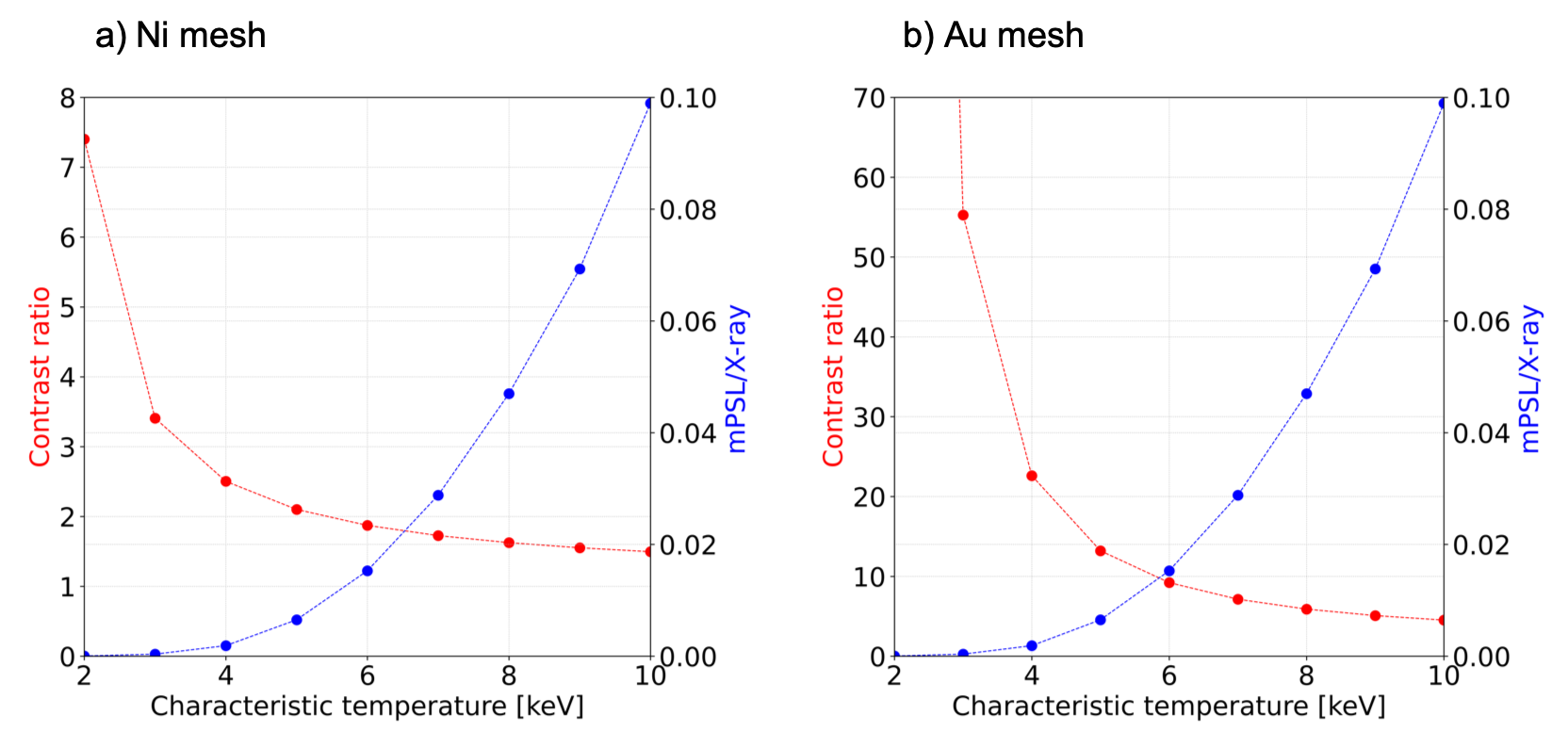}
\caption{The contrast ratio (red line) and signal level on the IP (blue line) as a function of the characteristic temperature of the initial X-ray spectrum for \textbf{a)} the Ni mesh and \textbf{b)} the Au mesh. }
\label{x_ray_contrastratio}
\end{figure}

The Fig. \ref{x_ray_contrastratio} shows the calculated contrast ratio of the X-ray image obtained on IP-MS as a function of the characteristic thermal temperature $T$ of the initial X-ray spectrum for a) a Ni mesh and b) a Au mesh. As one can see, the contrast ratio drastically increases as the temperature decreases, which is related to the increased absorption of lower-energy X-rays in the material. However, while contrast increases, one also needs to take into account the decrease of relative signal level on the IP due to the IP response function. We also can see that the use of a Au mesh instead of a Ni mesh would enhance X-ray contrast by a factor of 6, which is in overall agreement with FLUKA simulations.

\section{Conclusion}
In conclusion, we have demonstrated at the OMEGA and OMEGA-EP facilities a new design and implementation of proton radiography that includes an $\textit{in situ}$ X-ray reference.
The technique was validated by measuring non-uniform non-uniform vacuum magnetic fields generated by MIFEDS-powered coils. 
Importantly, the technique allows one to simultaneously obtain undeflected beamlet positions from the X-ray image of the mesh and deflected beamlet positions from the proton images, which is important for accurate calculation of the path-integrated magnetic field.
The success of this technique depends on high contrast in both the proton and X-ray images of the mesh so that the positions of the beamlets can be identified with high accuracy. In this work, we have used the FLUKA Monte-Carlo code to perform a comprehensive numerical study focused on the improvement of the technique using various mesh sizes and materials. We found that the best proton and X-ray image can be obtained by using a large Au mesh that provides enhanced contrast and reduced blurring.  Such enhancement of the contrast can significantly reduce the uncertainty in beamlets location and thereby improve the accuracy of the magnetic field measurement by approximately a factor of 2. We have also demonstrated for the first time the implementation of this technique using X-rays and TNSA protons generated by irritating a thin foil with a high-intensity laser.

\section{Backmatter}
\begin{backmatter}
\bmsection{Funding}
This work was supported through DOE Laboratory Directed Research and Development. The experiment was conducted at the Omega Laser Facility with the beam time through the Laboratory Basic Science under the auspices of the U.S. DOE/NNSA by the University of Rochester’s Laboratory for Laser Energetics under Contract DE-NA0003856. 

\bmsection{Acknowledgments}
We thank MIT team (A. Birkel, C.K.Li, P.J.Adrian, G. Sutcliffe) for supporting analysis of diagnostics. 
\bmsection{Disclosures}
The authors declare no conflicts of interest.

\bmsection{Data availability} Data underlying the results presented in this paper are not publicly available at this time but may be obtained from the authors upon reasonable request.

\end{backmatter}


\bibliography{sample}






\end{document}